# Observation of Giant Orbital Magnetic Moments and Paramagnetic Shift in Artificial Relativistic Atoms and Molecules

Zhehao Ge[1,*], Sergey Slizovskiy[2,3,*], Peter Polizogopoulos[1], Toyanath Joshi[1], Takashi Taniguchi[4], Kenji Watanabe[5], David Lederman[1], Vladimir I. Fal'ko[2,3,6, †], Jairo Velasco Jr.[1, †]

[1]*Department of Physics, University of California, Santa Cruz, California, USA*

[2]*Department of Physics and Astronomy, University of Manchester, Oxford Road, Manchester, M13 9PL, UK*

[3]*National Graphene Institute, University of Manchester, Booth Street East, Manchester, M13 9PL UK*

[4]*International Center for Materials Nanoarchitectronics National Institute for Materials Science, 1-1 Namiki, Tsukuba, 305-0044, Japan*

[5] *Research Center for Functional Materials National Institute for Materials Science, 1-1 Namiki, Tsukuba, 305-0044, Japan*

[6] *Henry Royce Institute for Advanced Materials, Manchester, M13 9PL, UK*

[*]These authors contributed equally to this manuscript.
[†]Email: jvelasc5@ucsc.edu, Vladimir.Falko@manchester.ac.uk

**Abstract:**

Massless Dirac fermions have been observed in various materials such as graphene and topological insulators in recent years, thus offering a solid-state platform to study relativistic quantum phenomena. Single quantum dots (QDs) and coupled QDs formed with massless Dirac fermions can be viewed as artificial relativistic atoms and molecules, respectively. Such structures offer a unique platform to study atomic and molecular physics in the ultra-relativistic regime. Here, we use a scanning tunneling microscope to create and probe single and coupled electrostatically defined graphene QDs to unravel the unique magnetic field responses of artificial relativistic nanostructures. Giant orbital Zeeman splitting and orbital magnetic moment up to ~70 meV/T and ~600$\mu_B$ are observed in single graphene QDs. While for coupled graphene QDs, Aharonov–Bohm oscillations and strong Van Vleck paramagnetic shift (~20 meV/T²) are observed. Such properties of artificial relativistic atoms and molecules can be leveraged for novel magnetic field sensing modalities.

Quantum dots (QDs) are often referred to as artificial atoms because of their atomic-like electronic structure[1,2]. They have been widely studied over the last 40 years in semiconductors and have provided immense fundamental insight[3-5]. Recently, the confinement of massless Dirac fermions in electrostatically defined QDs has been achieved in graphene[6-15] and topological insulators[16]. Different from semiconductor QDs formed with massive Schrödinger fermions, QDs populated by massless Dirac fermions can be viewed as artificial relativistic atoms, thus offering a unique opportunity to study atomic properties in the ultra-relativistic regime.

Graphene is an ideal platform for studying relativistic quantum phenomena because it hosts massless Dirac fermions[17] and has high tunability via electrostatic gating. As a result, multiple relativistic quantum phenomena have been demonstrated with graphene such as Klein tunneling[18,19] and atomic collapse[20,21]. Such phenomena are important not only for fundamental research but also for technological applications. For example, Klein tunneling renders graphene pn junctions highly transparent, which makes graphene an outstanding platform for electron optics applications such as negative refraction[22], Veselago lensing[23], and beam collimation[24,25].

When graphene massless Dirac fermions are confined into a quantum dot (QD), another intriguing platform for relativistic physics is realized, an artificial relativistic atom. For such a system, the usual relationship between orbital magnetic moment ($\vec{\mu}$) and angular momentum ($\vec{L}$) for atomic states ($\vec{\mu} = g\mu_B \vec{L}$, $\mu_B$ is the Bohr magneton) is invalid. This is because massless Dirac fermions disobey the non-relativistic relationship between velocity and momentum, $\vec{p} = m\vec{v}$. Instead, $\vec{\mu}$ is given by the area of the atomic orbit ($\pi r^2$) multiplied by the electrical current ($\frac{-ev_D}{2\pi r}$), which results in $\vec{\mu} = -e\vec{v}_D \times \vec{r}/2$. Because of this, the large and constant Dirac velocity $\vec{v}_D$ together with a sizable atomic orbital radius $\vec{r}$ can produce extremely large $\vec{\mu}$ for artificial relativistic atoms. One direct consequence of this large $\vec{\mu}$ is a giant Zeeman splitting for artificial atomic orbital states in a magnetic field ($B$), which can potentially be useful for sensing. Such properties of artificial relativistic atoms, however, have not been experimentally demonstrated to date. In this article we investigate single and coupled graphene QDs that are subjected to an external perpendicular $B$. These structures function as artificial relativistic atoms and molecules, and reveal intriguing $B$ responses that originate from the relativistic nature of these nanostructures such as giant orbital Zeeman splitting and strong paramagnetic shift.

## Observation of linear orbital Zeeman splitting

We study graphene QDs defined by electrostatically induced circular pn junctions with a scanning tunneling microscope (STM) as schematized in Fig. 1a. Although Klein tunneling[18,19] makes it difficult to confine massless Dirac fermions, their oblique incidence onto the circular pn junction boundary (schematized in Fig. 1b) avoids the 100% transmission occurring at normal incidence. This allows for the formation of quasi-bound states in graphene QDs, which have been confirmed in previous experiments[6,8-12,14,15]. In zero $B$, the clockwise and counterclockwise quasi-bound states possessing the same radial quantum number ($n$) and angular quantum numbers ($\pm m$) are degenerate due to time reversal symmetry. The directions of their $\vec{\mu}$, however, are opposite (Fig. 1c). Thus, by applying an external $B$, the degeneracy between the clockwise and counterclockwise quasi-bound states is lifted through an orbital Zeeman effect (Fig. 1d), leading to a splitting energy $\Delta E = 2|\vec{\mu} \cdot \vec{B}|$. This linear orbital Zeeman splitting can be used to measure $\vec{\mu}$ of graphene QD states.

Importantly, the Berry phase change of graphene QD states[10,26] in $B$ precludes the measurement of $\vec{\mu}$. To avoid this, we create graphene QDs with unprecedently sharp potential wells (further discussion in SI section S2). Our method involves using a two-step tip voltage pulsing technique based on prior works[9,27] (details in SI section S3) on samples with reduced hexagonal boron nitride (hBN) thickness. Figure 1e shows a typical $dI/dV_S$ spectra along a line across the center of a circular pn junction created with this technique on a large angle (14.1°) twisted bilayer graphene (tBLG)/hBN sample. By tracking the graphene charge neutrality point (marked by white dots in Fig. 1e), we estimate the potential variation to be 200~300 meV across 100 nm. This is 2~3 times sharper than previous works that utilized a related tip pulsing technique[9,10,14,15]. Figure 1f shows a comparison of $dI/dV_S$ point spectra at $d = 0$ nm and 40 nm of the graphene QD shown in Fig. 1e. Evidently, the $dI/dV_S$ peaks are much sharper off center than at the QD center. This is because near the QD boundary, states with larger $m$ are concentrated, which correspond to Dirac fermions propagating tangentially to the pn junction, resulting in a stronger reflection and hence better confinement[8,9]. For the remainder of this work, we will focus on these large $m$ states.

We now study the response of our graphene QDs to a perpendicular $B$. Figure 2a shows the comparison of $dI/dV_S(V_S, d)$ measured across the center of another graphene QD with a sharp potential well in $B = 0$ T and 0.2 T. Splitting patterns are clearly seen in $B = 0.2$ T as dimples near the QD boundary where high $m$ states concentrate. Figure 2b shows the evolution of $dI/dV_S$ point spectra at $d = 40$ nm in various $B$, the splitting and merging of graphene QD states can be seen as $B$ increases. To visualize this behavior more clearly, $dI/dV_S(V_S, B)$ with high $B$ resolution was acquired and is shown in Figure 2c. These data were taken from the same graphene QD shown in Fig. 2a at $d = 40$ nm, here $d^3I/dV_S^3(V_S, B)$ are presented to enhance the visibility of $dI/dV_S$ peaks (raw $dI/dV_S(V_S, B)$ data in Extended Data Fig. 1). We observe a clear linear splitting for each QD state. We attribute this behavior to orbital Zeeman splitting and find it is present at locations off the QD center but absent near the center where low m states concentrate (see SI section S4). These experimental findings are all in good agreement with simulations based on a tight binding (TB) model for a graphene QD (methods in SI section S5) with a quadratic potential well (Figs. 2d-f), thus further supporting our qualitative understanding. The slight deviations visible between experiment and simulation at negative energies is likely due to the deviation of the experimental potential from the quadratic potential used in our simulation.

**Angular quantum number and gate dependence of $\vec{\mu}$**

After confirming the existence of orbital Zeeman splitting, we now extract $\vec{\mu}$ of our artificial relativistic atom and study its angular quantum number and gate dependence. With spatially resolved $dI/dV_S$ spectra (Fig. 2a), we can assign both the radial and angular quantum numbers $(n, m)$ for graphene QD states in $d^3I/dV_S^3(V_S, B)$ plots[10] (details in SI section S6), the assigned quantum numbers are shown in Figs. 2c,f. By using the simple consideration discussed in Fig. 1d ($\Delta E = 2|\vec{\mu} \cdot \vec{B}|$), $\vec{\mu}$ of graphene QD states can then be extracted from the slopes of $\Delta E$ as a function of $B$. Figure 3a shows the extracted $\Delta E(B)$ and corresponding linear fits for QD states with different $m$ at $V_G = -16$ V (for more details see SI section S7), a clear increase in slope is seen for states with larger $m$. The magnitude of $\vec{\mu}$ ($\mu$) as a function of $m$ extracted from the slopes of linear fits in Fig. 3a is plotted in Fig. 3b. An increase from $\sim 200\mu_B$ to $\sim 500\mu_B$ for $\mu$ is seen when $m$ is increased from 2.5 to 10.5.

Next, we compare our experimentally extracted $\mu$ with theory. Approximately, the measured $\mu$ are on the order of $300\mu_B$, which agrees well with the $\mu$ of a current loop ($\mu = \frac{ev_F r}{2}$) for a charge flowing with graphene's Fermi velocity $v_F = 10^6$ m/s and with a loop radius $r = 35$ nm. For a more formal comparison we calculated $m$ resolved $LDOS(E, B)$ from a continuum model for graphene QDs with quadratic potential wells (details in SI section S8), some results are shown in Fig. 3c. From such plots, we can extract $\mu$ for graphene QD states with different quantum numbers, the extracted $\mu$ for states with $n = 0$ and $m = 2.5$ to $10.5$ are plotted in Fig. 3b. We notice the experimental results (red triangles) do not overlay any individual theory curve (empty circles). Additionally, experimental results at larger $m$ appear closer to theoretical curves calculated with a smaller $|\kappa|$. Akin to the discrepancy discussed for Fig. 2, the discrepancy seen here is likely caused by the deviation of the experimental potential from a quadratic potential at negative $V_S$ (further discussion in SI section S9). Nonetheless for both experiment and theory, a clear increase in $\mu$ is seen with increasing $m$.

We now explore the gate dependence of $\mu$ for our artificial relativistic atoms. Figure 3d shows $dI/dV_S(V_S, d)$ measured at $V_G = -20$ V, $-10$ V and 0 V in $B = 0.2$ T for the same QD in Fig. 2a. The potential well sharpness of graphene QDs in our experiments can be tuned by $V_G$ (further discussion in SI section S10). Apparently as the potential well sharpness increases with increasing $V_G$, the splitting energy reduces. Figure 3e shows the experimentally extracted $\Delta E(B)$ for the $n = 0, m = 5.5$ QD state at various $V_G$, $dI^3/dV_S^3(V_S, B)$ data at these $V_G$ are shown in Extended Data Fig. 2. A clear increase in slope is seen as $V_G$ is decreased, indicating an enhancement of $\mu$ when reducing $V_G$. To see this more quantitatively, we plot extracted $\mu$ values as a function of $m$ in Fig. 3f for graphene QD states with different $m$ measured at various $V_G$. The extracted $\mu$ are generally smaller at larger $V_G$ (sharper potential well) for all QD states.

The observed gate tuning of $\mu$ can be understood as a result of orbital size tuning of graphene QD states with $V_G$: sharper dots at larger $V_G$ have current loops with smaller radius. In contrast to nonrelativistic atoms, $\mu$ for relativistic atoms is governed by the orbital radius instead of the angular momentum. Therefore, it is uniquely possible to tune $\mu$ of our artificial relativistic atom by $V_G$ while maintaining the same quantum numbers. These results are also supported by the theoretically calculated $\mu(m)$ for graphene QDs with different potential well sharpness (Fig. 3b).

We now compare the observed $\mu$ in graphene QDs with that of other systems. The value of $\mu$ observed in this work are orders of magnitude larger than those observed in natural atoms[28] and semiconductor QDs[29-31], and are also several times larger than those observed in Bernal-stacked bilayer graphene (BLG) QDs[32,33] (Table 1). Although similar $\mu$ values have been observed in Bernal-stacked trilayer graphene (TLG) QDs[34], the findings in this work have several distinctions/advantages. Briefly, graphene QDs can achieve similar $\mu$ with a smaller QD size and maintain linear splitting within a larger $B$ range compared to TLG QDs (further discussion in SI section S11). Thus, the extremely large orbital Zeeman splitting observed in our graphene QDs (~23 to 58 meV/T) together with their nanometer scale sizes offers a unique opportunity to fabricate magnetometer arrays with nanometer scale spatial resolution (further discussion in SI section S12). This is difficult to achieve for the current state of the art[35,36].

## Coupled double graphene QDs

Having attained a thorough understanding of the $\mu$ of single graphene QDs, below we study the $B$ response of another interesting system, coupled double graphene QDs, which can be viewed as artificial relativistic molecules[37]. These structures were created on a graphene/hBN sample by fabricating two circular p-doped regions with centers separated by 150 nm with our two-step tip voltage pulsing technique (details in SI section S3). Figure 4a shows a $d^3I/dV_S^3(V_S, d)$ plot measured along a line across the centers of two dots in $B = 0$ T at $V_G = 0$ V, the red and blue patterns in the plot correspond to $dI/dV_S$ peaks and valleys, respectively. Three distinct regions can be identified and labeled as (i), (ii) and (iii) in Fig. 4a. At region (i), the energy spacing between $dI/dV_S$ peaks are half of those at regions (ii) and (iii); and at region (iii) different nodal patterns appear compared to region (ii). These features are distinct from uncoupled double graphene QDs made with a similar fabrication technique (see SI section S12).

Next, we map the $B$ response of our coupled graphene QDs. Plots of $d^3I/dV_S^3(V_S, B)$ measured at the three distinct regions of the coupled dots are shown in Figs. 4b and 4c. First, we notice at region (i) the QD states display a positive energy shift that is proportional to $B^2$, such behavior is more evident in a zoom in (right panel of Fig. 4b). The parabolic energy shift observed here is ~20 meV/T$^2$. Secondly, in region (ii) we encounter a linear splitting resembling the observation in single graphene QDs (left panel of Fig. 4c). Finally, region (iii) reveals a distinct behavior compared to the other regions, a staggered pattern for $dI/dV_S$ peaks (right panel of Fig. 4c). These observations are all qualitatively reproduced with a TB model for a coupled double graphene QD (see SI section S13).

To attain an intuitive understanding of this rich set of observations, we consider semiclassical orbits within coupled double graphene QDs. For QD states in a single graphene QD, their semiclassical orbits can be approximated as circular orbits. Once two QD states couple, the circular orbits of individual QD states merge into a figure-eight orbit. Such an orbit is schematized by the yellow rings embedded in a double graphene QD in Fig. 4d. The arrows indicate the direction of the current, a degenerate time-reversed figure-eight orbit also exists but is not shown here for clarity. Closer to each individual QD center, we expect QD states have a smaller radius and are decoupled thus forming circular orbits, as depicted by green rings in Fig. 4d. In our experiment, regions (i) and (iii) correspond to QD states with figure-eight orbits, and region (ii) corresponds to QD states with circular orbits. With this understanding in hand, the half energy spacing observed at region (i) compared to region (ii) is due to the length of the figure-eight orbit being twice that of the circular orbit. This is because according to the semiclassical quantization rule the energy spacing between graphene QD states is $\Delta E = \frac{hv_F}{L}$, where h is Planck's constant, $v_F$ is the graphene Fermi velocity and $L$ is the semiclassical orbit length. Consequently, the large $B$ induced linear splitting observed at region (ii) can be explained by the large $\vec{\mu}$ of circular orbits akin to uncoupled graphene QDs.

**Van Vleck paramagnetic shift and Aharonov–Bohm effect**

We now discuss how the unique $B$ response observed in region (i) corresponds to the emergence of novel magnetism due to the relativistic nature of our artificial molecule. Because each of the two rings of the figure-eight orbit have reversed current flow directions, their $\vec{\mu}$ are in opposite directions but with the same magnitude. Therefore, the net $\vec{\mu}$ of an entire figure-eight orbit will be zero, hence explaining the disappearance of linear splitting in $B$ at regions (i) and (iii). Moreover, the positive parabolic energy shift for holes observed at region (i) is caused by a Van Vleck paramagnetic shift, which is a second order perturbative $B$ response[38]. Classically, this effect stems from a Lorentz force that expands (contracts) orbits with $\mu$ aligned (anti-aligned) to $B$, resulting in an increase (decrease) of $\mu$. The quenching of the first order $B$ effect in figure-eight orbits helps the detection of this second order effect. Importantly, for non-relativistic systems such as natural atoms[39] and semiconductor QDs[29,30,39], a Larmor diamagnetic shift due to the change of electron orbital velocity in $B$ also exists and is usually stronger than the Van Vleck paramagnetic shift. However, for graphene QDs, the Larmor diamagnetic shift is absent because of the constant velocity of massless Dirac fermions. Alternatively, this can be understood as resulting from the graphene Hamiltonian in $B$ lacking a $B^2$ term that produces Larmor diamagnetism in non-relativistic systems (further discussion in SI section S14). Such a $B$ induced response is thus unique to ultra-relativistic artificial atoms and molecules.

Finally, we discuss the origin of the staggered red stripe patterns in $dI^3/dV_S^3(V_S, B)$ plot at region (iii). We attribute this phenomenon to Aharonov–Bohm (AB) oscillations that occur at the crossing point of figure-eight orbits, region (iii). When a particle returns to its original position after traveling along a closed path, constructive (destructive) interference leading to enhanced (reduced) $LDOS$ will occur if the action along the closed path $\frac{1}{\hbar}\oint \boldsymbol{p} \cdot d\boldsymbol{q}$ equals an even (odd) number of $\pi$. Here $\boldsymbol{p}$ and $\boldsymbol{q}$ are the canonical momentum and position coordinates, $\hbar$ is Dirac's constant. As shown in Fig. 4e, two distinct eigen energies exist for figure-eight orbits with constructive (yellow dot) and destructive (black dot) interference at the figure-eight orbit center, respectively. This is because charges return to this location after traveling half of a figure-eight orbit. By applying an external $B$, the AB effect causes the pickup of additional phases ($\Delta\varphi = \frac{-e\Phi_B}{\hbar}$, where $\Phi_B$ is the magnetic flux through each circular segment) with opposite signs for the two circular orbits flowing in opposite directions. As a result, the energy level of the figure-eight orbit does not change in $B$ to first order, but the interference type at the figure-eight orbit center switches depending on the amount of $\Delta\varphi$ picked up by each circular segment (Fig. 4e). This explains the staggered pattern observed at region (iii) (Fig. 4c), where red and blue stripes (corresponding to constructive and destructive interference, respectively) alternate at a constant $V_S$ in $B$. In addition, the assigned $\Delta\varphi$ in Fig. 4c are in good agreement with our experimental double dot geometry (see SI section S15), thus further supporting our interpretation. This observed AB oscillation in the $LDOS$ intensity of coupled graphene QDs can also be potentially used for $B$ sensing. Notably, it represents a different modality compared to the linear orbital Zeeman splitting observed in single graphene QDs.

## Conclusions

In conclusion, we observed giant $\vec{\mu}$ and large orbital Zeeman splitting in artificial relativistic atoms formed with single graphene QDs. We also observed strong Van Vleck paramagnetic shift and AB oscillations in artificial relativistic molecules formed with coupled double graphene QDs. Our work adds fundamental insight into relativistic quantum phenomena in solid state systems and paves the way towards new modalities of $B$ sensing that utilizes massless Dirac fermions.

## Methods

**Sample fabrication.** The graphene/hBN and tBLG/hBN samples were assembled with a standard polymer-based transfer method[40]. For the graphene/hBN sample, a graphene flake exfoliated on a methyl methacrylate (MMA) substrate was mechanically placed on top of an ∼20 nm thick hBN flake that rests on a 285 nm $SiO_2/Si^{++}$ substrate. For the tBLG sample, we first use a monolayer graphene flake on MMA to pick up another monolayer graphene flake on $SiO_2/Si^{++}$ substrate, then we place the tBLG/MMA onto an ∼20 nm hBN flake that rests on a 285 nm $SiO_2/Si^{++}$ substrate. For both samples, the MMA scaffold was dissolved in a subsequent solvent bath. The assembled heterostructure is then annealed in forming gas ($Ar/H_2$) for ∼12 hours at 400 °C to reduce residual polymer after the heterostructure assembly procedure. Next, an electrical contact to the sample is made by thermally evaporating 7 nm of Cr and 200 nm of Au using a metallic stencil mask. To further improve the sample surface cleanliness, the heterostructure is then mechanically cleaned using an AFM[41], which is done in a glovebox filled with $N_2$ gas. We perform sequential scans in contact mode (setpoint of 0.2 V, scanning speed of ∼15 μm/sec, and $1024 \times 1024$ pixels resolution) to sweep regions of ∼ $30 \times 30$ μm² by a Cypher S AFM with Econo-ESP-Au tips from Oxford Instruments. Finally, the heterostructure is annealed in ultra-high vacuum (UHV) at 400 °C for seven hours before being introduced into the STM chamber.

**STM/STS measurements.** The STM/STS measurements were conducted in UHV with pressures better than $1 \times 10^{-10}$ mbar at 4.8 K in a Createc LT-STM. Electrochemically etched tungsten tips calibrated on Au(111) surface were used in the experiments. The lock-in AC signal frequency used for STS measurements was 704 Hz.

**Acknowledgments:** We thank the Hummingbird Computational Cluster team at UC Santa Cruz for providing computational resources for the numerical TB calculations performed in this work, and Mike Hance for providing insight on accelerator physics considerations related to our experimental findings. J.V.J. and Z.G. acknowledges support from the National Science Foundation under award DMR-1753367. J.V.J acknowledges support from the Army Research Office under contract W911NF-17-1-0473. V.I.F. and S.S. acknowledge support from the European Graphene Flagship Core 3 Project. V.I.F. acknowledges support from Lloyd Register Foundation Nanotechnology Grant, EPSRC grants EP/V007033/1, EP/S030719/1 and EP/N010345/1. K.W. and T.T. acknowledge support from the Elemental Strategy Initiative conducted by the MEXT, Japan, Grant Number JPMXP0112101001, JSPS KAKENHI Grant Numbers JP20H00354.

**Author contributions:** Z.G. and J.V.J. conceived the work and designed the research strategy. Z.G. fabricated the samples and performed data analysis under J.V.J.'s supervision. K.W. and T.T. provided the hBN crystals. Z.G. carried out tunneling spectroscopy measurements with assistance from P.P. and T.J. under D.L. and J.V.J.'s supervision. S.S. developed the interpretation for experimental findings and performed continuum model calculations under V.I.F.'s supervision. Z.G. performed numerical TB calculations with input from S.S. under V.I.F. and J.VJ.'s supervision. Z.G. and J.V.J. wrote the paper. All authors discussed the paper and commented on the manuscript.

# Figure 1

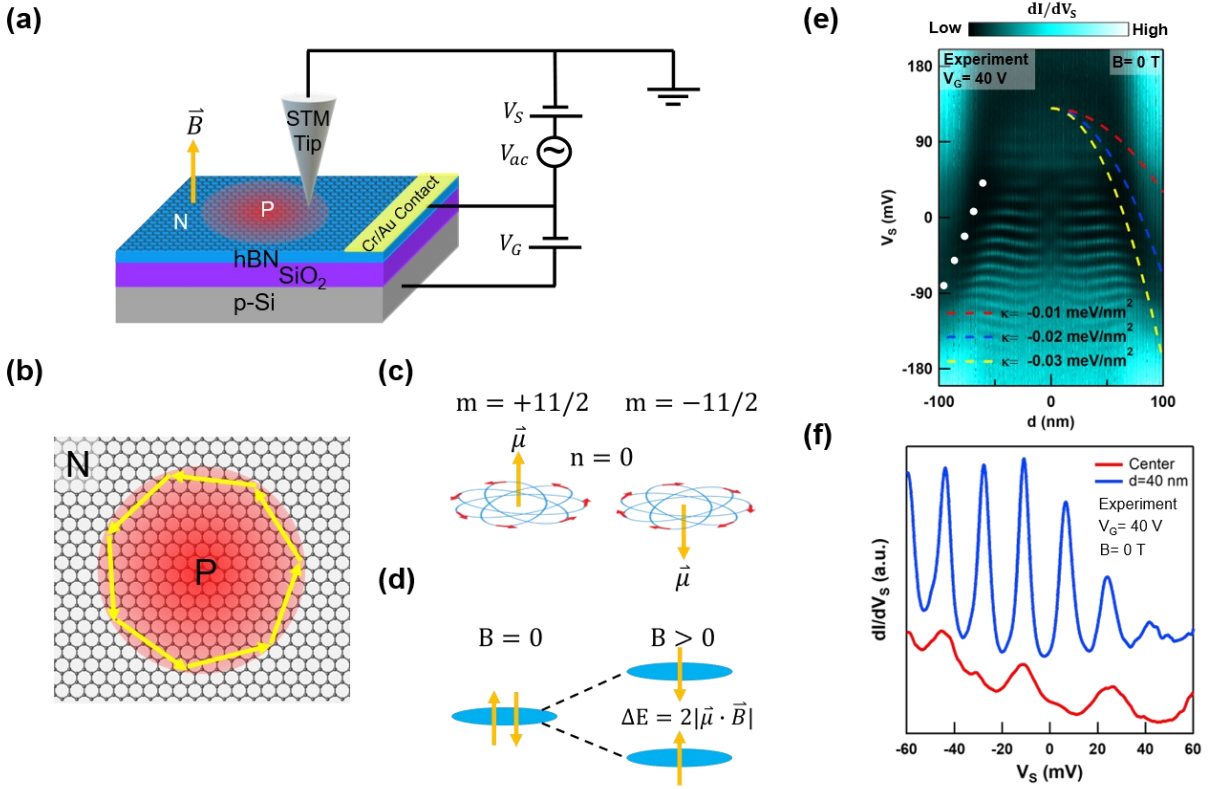

**Figure 1: Experimental set up and orbital Zeeman splitting of graphene Quantum Dot (QD) states. a,** Schematic of the experimental setup. The circular graphene pn junction is created in a monolayer graphene or large angle twisted bilayer graphene/hBN heterostructure (additional information in SI section S1) resting on a SiO$_2$/Si chip. Contact to graphene is made through a Cr/Au electrode. The STM tip is grounded, a bias voltage $V_S$ together with an ac voltage $V_{ac}$ is applied between the STM tip and graphene. A backgate voltage $V_G$ is applied between the p-doped silicon and graphene and an out of plane magnetic field is applied to the whole device. **b,** Schematic of the confinement of massless Dirac fermions in a circular graphene pn junction. **c,** Schematic of the orbital magnetic moments of graphene QD states. The blue lines are the calculated semiclassical orbits of $n = 0$, $m = \pm 11/2$ graphene QD states in a parabolic potential well $U(r) = \kappa r^2$, where $\kappa = -0.01$ meV/nm$^2$. The red arrows indicate the direction of the trajectory. The orange arrows indicate the orientation of the orbital magnetic moments. **d,** Schematic of the orbital Zeeman splitting of graphene QD states in a finite magnetic field. The blue ovals and orange arrows represent the energy levels and the orbital magnetic moment orientations of graphene QD states, respectively. **e,** Experimentally measured $dI/dV_S(V_S, d)$ at $V_G = 0$ V along a line across the center of a circular graphene pn junction that has a sharp potential well. Colored dashed lines are quadratic potential wells with different $\kappa$ values. The set point used to acquire the tunneling spectra was $I = 1$ nA, $V_S = -200$ mV, with a 2 mV ac modulation. **f,** $dI/dV_S$ point spectra at the center and at 40 nm away from the center of the circular graphene pn junction as shown in (c). The set point used to acquire the tunneling spectra was $I = 1$ nA, $V_S = -60$ mV, with a 2 mV ac modulation.

# Figure 2

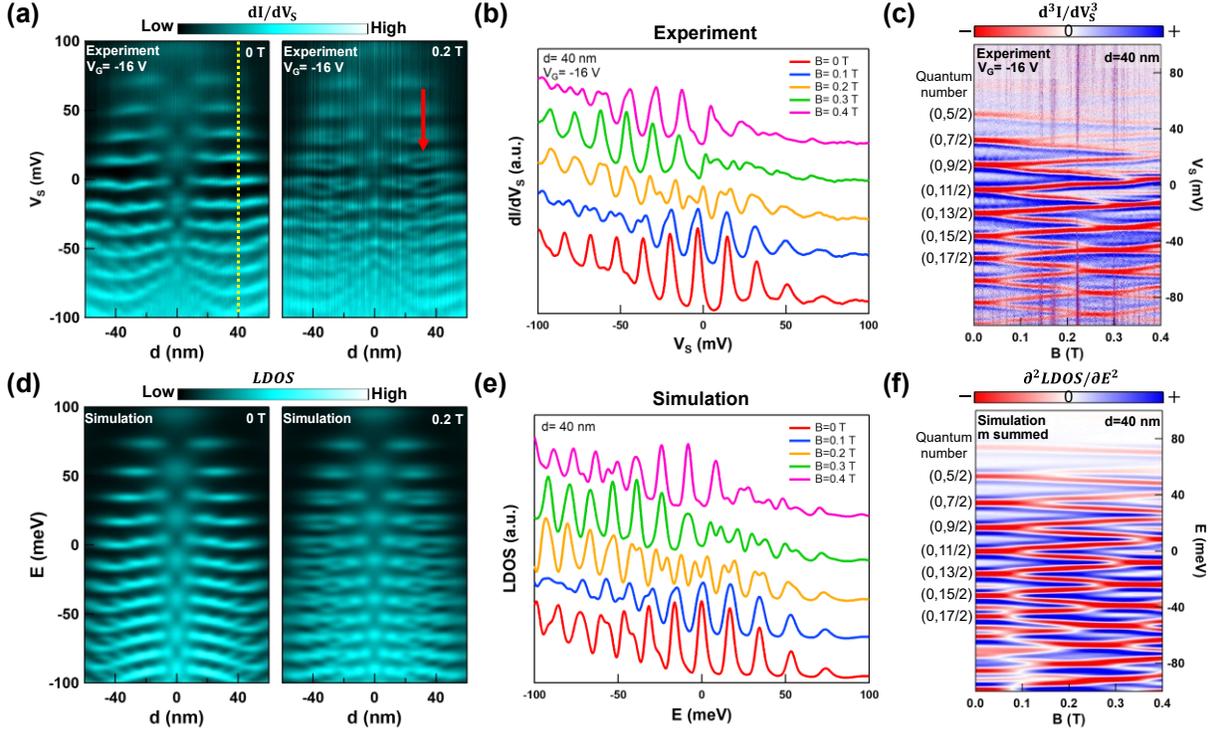

**Figure 2: Experimental observation of linear orbital Zeeman splitting. a,** Experimentally measured $dI/dV_S(V_S, d)$ at $V_G = -16$ V in $B = 0$ T and $B = 0.2$ T along a line across the center of a circular graphene pn junction with a sharp potential well. This pn junction is different from the junction shown in Fig. 1e. The red arrow indicates the splitting of one QD state. **b,** $dI/dV_S$ point spectra measured at $d = 40$ nm and at $V_G = -16$ V in various magnetic fields from 0 T to 0.4 T with a 0.1 T step. **c,** $d^3I/dV_S^3(V_S, B)$ at $V_G = -16$ V at $d = 40$ nm as indicated by the yellow dashed lines in (a). The quantum number $(n, m)$ corresponds to radial and angular quantum number, respectively. **d,** Simulated $LDOS(E, d)$ for a graphene QD in $B = 0$ T and $B = 0.2$ T with $U(r) = -0.03r^2$ meV/nm$^2$ + 160 meV. **e,** Simulated $LDOS$ at $d = 40$ nm in various magnetic fields from 0 T to 0.4 T for the same graphene QD in (d). **f,** Simulated $\partial^2 LDOS/\partial E^2(E, B)$ at $d = 40$ nm for the same graphene QD in (d). The quantum number $(n, m)$ corresponds to radial and angular quantum number, respectively.

**Figure 3**

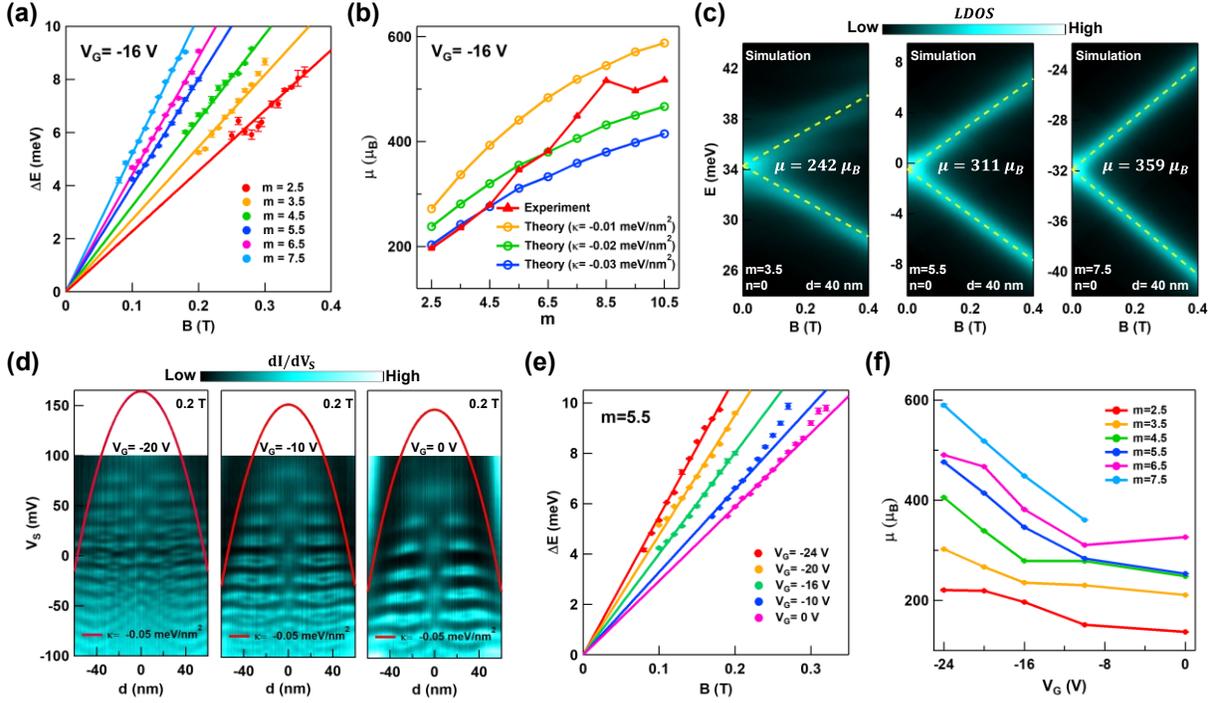

**Figure 3: Quantum number and gate dependence of magnetic moments of graphene QD states. a,** Experimentally extracted orbital Zeeman splitting energy at various $B$ and linear fits for graphene QD states with different $m$ and $n = 0$ at $V_G = -16$ V. **b,** Comparison between experimentally extracted $\mu(m)$ at $V_G = -16$ V and theoretically calculated $\mu(m)$ for graphene QD states with parabolic potential wells $U(r) = \kappa r^2$. **c,** Calculated $m$ resolved $LDOS(E, B)$ at $d = 40$ nm for a graphene QD with $U(r) = -0.03 r^2$ meV/nm$^2$ + 160 meV. Yellow dashed lines indicate the theoretical orbital Zeeman splitting size with the $\mu$ given in each plot. **d,** Experimentally measured $dI/dV_S(V_S, d)$ at $V_G = -20$ V, $-10$ V and 0 V in $B = 0.2$ T along a line across the center of the same circular graphene pn junction shown in Fig. 2a. The red curve in each plot represents a parabolic potential well with $\kappa = -0.05$ meV/nm$^2$ and is a guide for the eye to aid comparison between potential well sharpness variation between different $V_G$. **e,** Experimentally extracted orbital Zeeman splitting energy at various $B$ and corresponding linear fits for graphene QD states with $n = 0, m = 5.5$ at different $V_G$. **f,** Experimentally extracted $\mu$ for graphene QD states with different $m$ and $n = 0$ at different $V_G$. Error bars in (a) and (e) reflect the uncertainty of the peak position in Gaussian fitting. Error bars in (b) and (f) reflect the uncertainty of the slope in weighted linear fitting.

# Figure 4

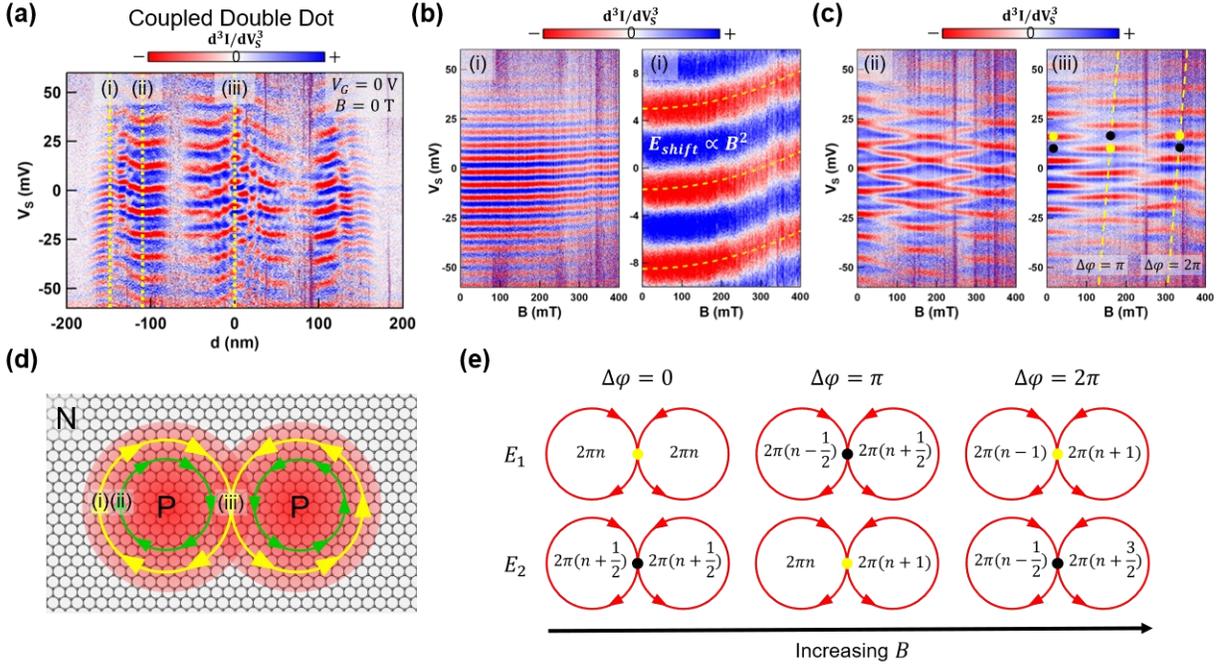

**Figure 4: Paramagnetic shift and Aharonov–Bohm (AB) effect in coupled double graphene QDs. a,** Experimentally measured $d^3I/dV_S^3(V_S, d)$ at $V_G = 0$ V in $B = 0$ T for coupled double graphene QDs that are separated by 150 nm. **b,** Left panel: $d^3I/dV_S^3(V_S, B)$ measured at $V_G = 0$ V at $d = -150$ nm corresponding to region (i) in (a). Right panel: zoom in around $V_S = 0$ of the left panel. **c,** Left panel: $d^3I/dV_S^3(V_S, B)$ measured at $V_G = 0$ V at $d = -110$ nm corresponding to region (ii) in (a). Right panel: $d^3I/dV_S^3(V_S, B)$ measured at $V_G = 0$ V at $d = 0$ nm corresponding to region (iii) in (a). Yellow dashed lines indicate approximate $B$, at which the circular orbits pick up an integer number of $\pi$ AB phase. $d^3I/dV_S^3$ values in (a-c) were numerically calculated from the $dI/dV_S$ acquired through a lock-in measurement and smoothed with a 2 mV box average. The set point used to acquire the tunneling spectra in (a-c) was $I = 1$ nA, $V_S = -60$ mV, with a 2 mV ac modulation. **d,** Schematic of coupled double graphene QD and figure-eight and circular orbits. **e,** Schematic of the constructive (represented by yellow dot) and destructive (represented by black dot) interference at the center of the figure-eight orbit for different energy levels and its tuning by $B$ through AB effect.

# Table 1

| Systems | Observed maximum effective $\mu$ |
|---|---|
| Natural Atoms | On the order of several $\mu_B$ for their ground states and not highly excited states[28] |
| Self-assembled InP QDs[29] | $\sim 17.3\mu_B$ ($\sim 2.0$ meV/T) |
| Self-assembled InAs/GaAs QDs[30] | $\sim 18.5\mu_B$ ($\sim 2.1$ meV/T) |
| Self-assembled InGaAs/GaAs QDs[31] | $\sim 19.5\mu_B$ ($\sim 2.3$ meV/T) |
| STM tip induced BLG QDs[32] | $\sim 86\mu_B$ ($\sim 10$ meV/T) |
| Few-carrier BLG QDs[33] | $\sim 45\mu_B$ ($\sim 5$ meV/T) |
| STM tip induced TLG QDs[34] | $\sim 525\mu_B$ ($\sim 61$ meV/T) |
| This work (MLG QDs) | $\sim 600\mu_B$ ($\sim 70$ meV/T) |

**Table 1: Comparison of measured $\mu$ values from different types of systems.** For works in which $\mu$ values are not given directly, we convert the observed Zeeman splitting $\Delta E$ in $B$ to an effective $\mu$ through the definition $\mu = \frac{\Delta E}{2B}$.

# Extended Data Figure 1

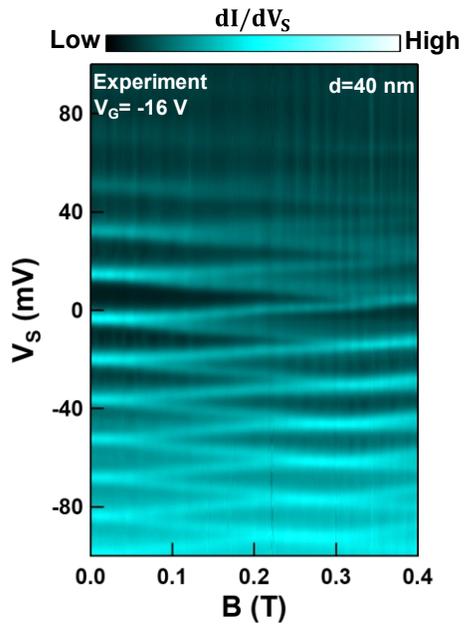

**Extended Data Figure 1: Raw $dI/dV_S(V_S, B)$ used to get $d^3I/dV_S^3(V_S, B)$ plot in Fig. 2c.**

# Extended Data Figure 2

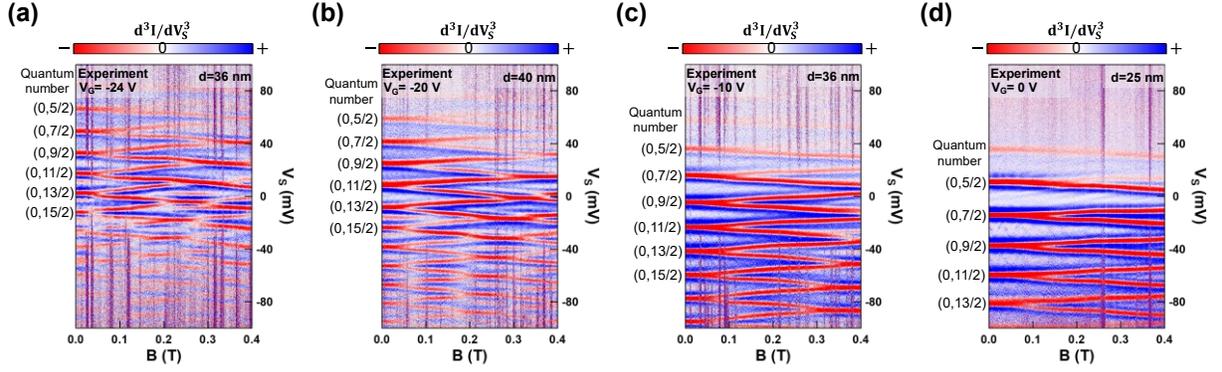

**Extended Data Figure 2:** $d^3I/dV_S^3(V_S, B)$ **plot at different** $V_G$. **a,** $d^3I/dV_S^3(V_S, B)$ at $V_G = -24$ V and at $d = 36$ nm. **b,** $d^3I/dV_S^3(V_S, B)$ at $V_G = -20$ V and at $d = 40$ nm. **c,** $d^3I/dV_S^3(V_S, B)$ at $V_G = -10$ V and at $d = 36$ nm. **d,** $d^3I/dV_S^3(V_S, B)$ at $V_G = 0$ V and at $d = 25$ nm. The quantum number $(n, m)$ in (a-d) corresponds to radial and angular quantum number, respectively.

12	Bai, K.-K. *et al.* Generating atomically sharp p− n junctions in graphene and testing quantum electron optics on the nanoscale. *Physical Review B* **97**, 045413 (2018).

13	Freitag, N. M. *et al.* Large tunable valley splitting in edge-free graphene quantum dots on boron nitride. *Nature nanotechnology* **13**, 392-397 (2018).

14	Quezada-López, E. A. *et al.* Comprehensive Electrostatic Modeling of Exposed Quantum Dots in Graphene/Hexagonal Boron Nitride Heterostructures. *Nanomaterials* **10**, 1154 (2020).

15	Behn, W. A. *et al.* Measuring and Tuning the Potential Landscape of Electrostatically Defined Quantum Dots in Graphene. *Nano Letters* **21**, 5013-5020 (2021).

16	Zhang, J., Jiang, Y.-P., Ma, X.-C. & Xue, Q.-K. Berry-Phase Switch in Electrostatically Confined Topological Surface States. *Physical Review Letters* **128**, 126402 (2022).

17	Novoselov, K. S. *et al.* Two-dimensional gas of massless Dirac fermions in graphene. *nature* **438**, 197-200 (2005).

18	Katsnelson, M., Novoselov, K. & Geim, A. Chiral tunnelling and the Klein paradox in graphene. *Nature physics* **2**, 620-625 (2006).

19	Young, A. F. & Kim, P. Quantum interference and Klein tunnelling in graphene heterojunctions. *Nature Physics* **5**, 222-226 (2009).

20	Wang, Y. *et al.* Observing atomic collapse resonances in artificial nuclei on graphene. *Science* **340**, 734-737 (2013).

21	Lu, J. *et al.* Frustrated supercritical collapse in tunable charge arrays on graphene. *Nature communications* **10**, 1-8 (2019).

22	Chen, S. *et al.* Electron optics with pn junctions in ballistic graphene. *Science* **353**, 1522-1525 (2016).
22